\documentclass[conference]{IEEEtran}
\IEEEoverridecommandlockouts
\usepackage{cite}
\usepackage{amsmath,amssymb,amsfonts}
\usepackage{algorithmic}
\usepackage{algorithm}  
\usepackage{graphicx}
\usepackage{textcomp}
\usepackage{xcolor}
\usepackage{float} 
\usepackage{bm}
\usepackage{multirow}
\usepackage{dblfloatfix} 
\usepackage{float} 
\usepackage{array}
\usepackage{booktabs}
\usepackage{multirow}
\usepackage{fancyhdr} 
\usepackage{color}
\usepackage{tabu}
\usepackage{nomencl}
\makenomenclature
\usepackage{etoolbox}
\usepackage{graphicx}
\usepackage{subfigure}
\usepackage{amsfonts,amssymb}
\usepackage[cal=boondoxo]{mathalfa}
\usepackage[colorlinks, citecolor=black]{hyperref}
\hypersetup{ colorlinks=true, linkcolor=black, citecolor=black, urlcolor = black }
\newcolumntype{I}{!{\vrule width 1.3pt}}
\newlength\savedwidth

\newlength\savewidth

\def\BibTeX{{\rm B\kern-.05em{\sc i\kern-.025em b}\kern-.08em
    T\kern-.1667em\lower.7ex\hbox{E}\kern-.125emX}}

\IEEEoverridecommandlockouts\IEEEpubid{\makebox[\columnwidth]{ 978-1-6654-3540-6/22~\copyright~2022 IEEE \hfill} \hspace{\columnsep}\makebox[\columnwidth]{ }}

\begin{document}

\title{Hierarchical Reinforcement Learning for RIS-Assisted Energy-Efficient RAN
\\
}

\author{\IEEEauthorblockN{Hao Zhou$^1$, Long Kong$^1$, Medhat Elsayed$^2$, Majid Bavand$^2$, Raimundas Gaigalas$^3$, \\ 
Steve Furr$^2$, and Melike Erol-Kantarci$^1$, \IEEEmembership{Senior Member, IEEE}}
\IEEEauthorblockA{{$^1$School of Electrical Engineering and Computer Science, University of Ottawa, Ottawa, Ontario, Canada} \\
{$^2$ Ericsson Canada, Ottawa, Ontario, Canada} \qquad
{$^3$ Ericsson Sweden, Stockholm County, Sweden}\\
Emails: \{hzhou98, lkong2, melike.erolkantarci\}@uottawa.ca}\{medhat.elsayed,majid.bavand,raimundas.gaigalas,steve.furr\}@ericsson.com\\
}

\maketitle

\thispagestyle{fancy} %
      \lhead{} 
      \chead{Accepted by 2022 IEEE Globecom conference, \copyright2022 IEEE } 
      \rhead{} 
      \lfoot{} 
      \cfoot{\thepage} 
      \rfoot{} 
      \renewcommand{\headrulewidth}{0pt} 
      \renewcommand{\footrulewidth}{0pt} 
\pagestyle{fancy}

\begin{abstract}
Reconfigurable intelligent surface (RIS) is emerging as a promising technology to boost the energy efficiency (EE) of 5G beyond and 6G networks. Inspired by this potential, in this paper, we investigate the RIS-assisted energy-efficient radio access networks (RAN). In particular, we combine RIS with sleep control techniques, and develop a hierarchical reinforcement learning (HRL) algorithm for network management. In HRL, the meta-controller decides the on/off status of the small base stations (SBSs) in heterogeneous networks, while the sub-controller can change the transmission power levels of SBSs to save energy. The simulations show that the RIS-assisted sleep control can achieve significantly lower energy consumption, higher throughput, and more than doubled energy efficiency than no-RIS conditions.
\end{abstract}


\section{Introduction}
In line with previous generations of mobile wireless technologies, 5G is currently on the road to mass deployment. Meanwhile, the energy efficiency of 5G has been a significant research area in academia and industry \cite{b8}.  As stated in \cite{7041163}, one of the widely considered approaches for energy efficiency has been the sleep control technique.
Sleep control refers to selectively turning radio transceivers or base stations (BSs) to sleep mode. Different than 4G, recurring transmission of always-on signals to guarantee network coverage, the 5G new radio (NR) standard allows to deploy the sleep mode. 

More recently, reconfigurable intelligent surfaces (RISs) are proposed and considered as key enablers for future wireless communications \cite{9140329}. RIS is essentially an electronically operated metasurface controlled by programmable software, which is physically equivalent to digitally controllable scatterers and software defined
surface \cite{9473681}. A large number of small, low-cost, and passive artificial “meta-atoms” integrated into the RIS can smartly change the reflection direction towards any desired users by tuning a series of phase shifters. Accordingly, RISs have been designed for various scenarios and applications including 6G, internet of things (IoT), smart cities \cite{9253607}, etc. The main benefit of RIS lies in its capability of shaping the wireless propagation environments by adjusting the signal reflections \cite{9140329}. Through this, the signal quality and connectivity can be substantially improved. Furthermore, the energy consumption of RIS is extremely low, which is a favourable property compared to traditional relaying \cite{8888223}. RIS's capability and low-power consumption features motivate us to investigate the RIS-aided energy-efficient RAN.

Moreover, machine learning has been generally applied for wireless network management for its advantage in handling dynamic environment\cite{b1}. For example, in reinforcement learning (RL), the optimization problem can be transformed to the unified Markov decision process (MDPs), which avoids the complexity of defining a dedicated optimization model. In this paper, the main contribution is that we propose a novel hierarchical reinforcement learning (HRL) architecture for RIS-assisted sleep control in heterogeneous networks. 
Compared with conventional RL that includes one standalone agent, HRL defines a meta-controller and a sub-controller, which enables higher exploration efficiency by the hierarchical architecture \cite{bsurvey}.
In particular, we improve the energy efficiency (EE) in two ways: we consider the macro base station (MBS) as the meta-controller to implement the sleep control of small base stations (SBSs) to save energy, and SBSs as sub-controllers to decide its own transmission power level to reduce energy consumption. Besides, RIS is deployed to improve the signal propagation environment and increase the channel capacity. Finally, the simulations show that combining RIS with sleep control can achieve lower energy consumption, higher throughput, and more than doubled EE than the standalone sleep control strategy.  

\section{Related work}
\subsection{Machine learning based sleep control}
The flourishing machine learning techniques offer promising opportunities for network control and management \cite{b9}. The deep Q-network is deployed in \cite{b2} for the sleep control of renewable energy-powered BSs, where the SBSs can share their energy by a micro-grid. The neural network is applied in \cite{8741057} to predict both traffic demand and energy production, and then the prediction results are used for sleep control of BSs. Similarly, \cite{b3} deploys deep neural networks to predict the traffic patterns, and actor-critic reinforcement learning is used for dynamic sleep control. Different from aforementioned works, here we apply the HRL algorithm, including a meta-controller for SBS sleep control and sub-controllers for the transmission power control. This hierarchical control strategy allows more efficient exploration of the environment, and mitigates the long convergence issue of conventional RL\cite{bsurvey}.           

\subsection{RIS related researches}
RIS, being an appealing approach, has become a hot topic for researchers from both wireless communication and signal processing communities \cite{b11}. The machine learning-enabled RIS-assisted wireless communication systems have been under exploration in terms of channel modelling \cite{9312154,9473681}, channel estimation, EE \cite{9149380}, etc. More specifically, the authors in \cite{9473681} and \cite{9312154} applied the unsupervised machine learning tool, namely, the expectation-maximization (EM) algorithm to model the RIS-assisted wireless communication links. It is also demonstrated that the machine learning methods are able to provide better performance than central limit theorem-based approaches \cite{9473681} \cite{9312154}. Besides, Lee \textit{et al}. in \cite{9149380} deployed the deep reinforcement learning to improve the EE of the RIS-aided cellular communication systems, but the sleep control is not involved. 
To the best knowledge of the authors, no work before has ever investigated the EE problem with sleep control and RIS embedded in the cellular communication systems. 

\section{Network and System Model}
\begin{figure}[!t]
\centering
\includegraphics[width=0.75\linewidth]{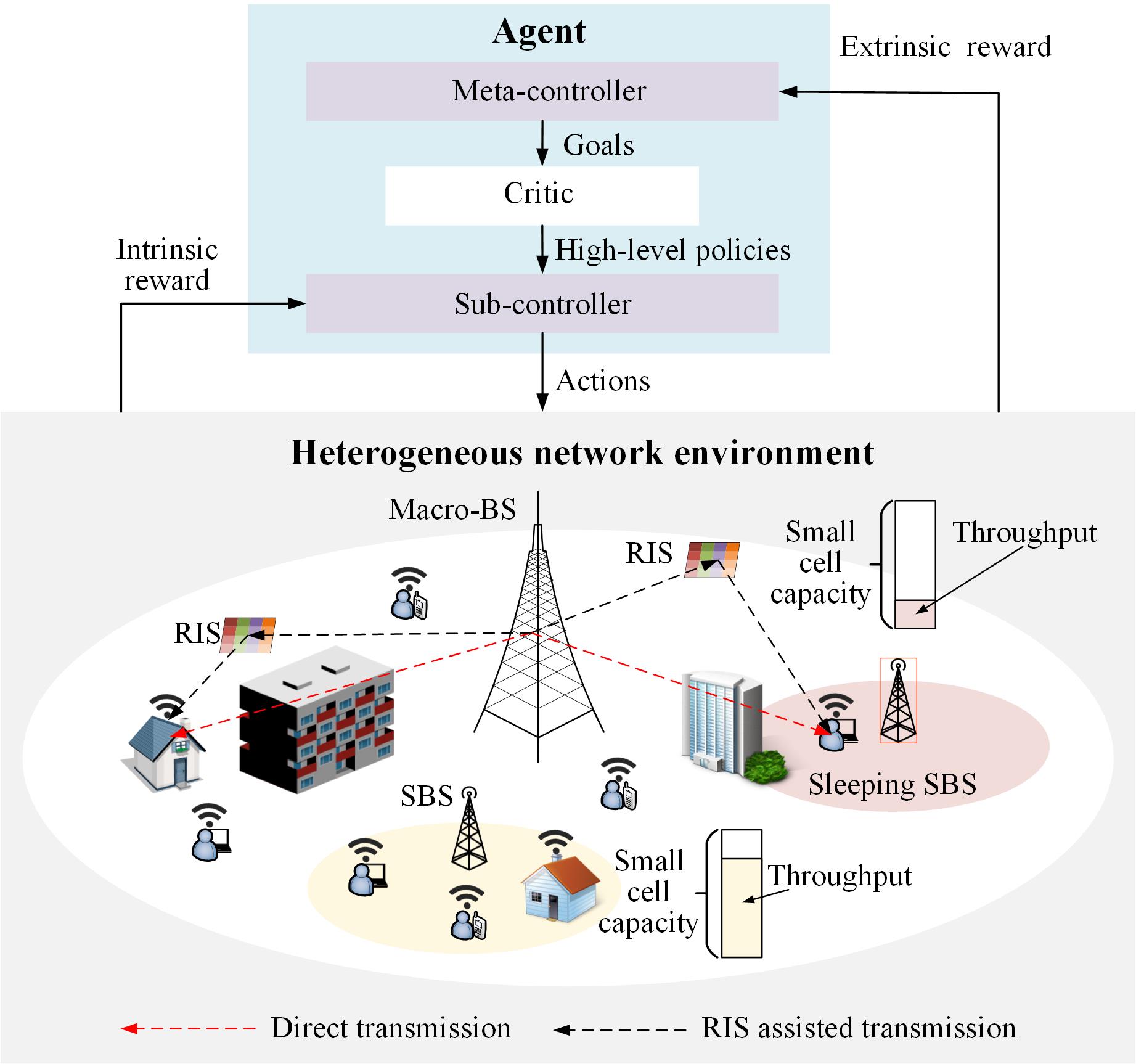}
\caption{RIS-aided heterogeneous network.}
\label{fig1}
\setlength{\abovecaptionskip}{-2pt} 
\vspace{-15pt}
\end{figure}

As illustrated in Fig.\ref{fig1}, we consider a heterogeneous network that includes one MBS and several SBSs. The SBS may switch to the sleep mode when the traffic demand drops, which will reduce the energy consumption. It is assumed the MBS can take over the active user equipment (UEs) that were previously associated with those small cells. On the other hand, high-density buildings in the urban area lead to high penetration loss and lower received signal-to-interference-plus-noise ratio (SINR) for direct transmissions. To this end, we deploy RIS to reflect the signal from MBS and mitigate the high penetration loss of direct transmissions. In this work, it is worth noting that we save energy in two ways: (i) the meta-controller decides the on/off status of SBSs when traffic load changes, and (ii) the sub-controller decides the transmission power of active SBSs. This hierarchical architecture enables a higher management efficiency, which will be introduced in detail in Section \ref{s4}. 

\subsection{RIS-Assisted Channel Model}
It is assumed that UEs can receive the signal from BSs by direct and indirect transmissions. The direct link is considered as non-line-of-sight (NLOS) transmission due to the dense buildings in the urban area. The baseband equivalent channel between BS-UE is given by: 
\begin{equation}
 \label{eq1}
\bm{H}_{Bk}=g_{B,k}\bm{h}_{B,k},
\end{equation}
where $g_{B,k}$ is the path loss from BS to UE, and $\bm{h}_{B,k}$ is a complex Gaussian distributed random vector with 0 mean and unit variance, i.e., $\mathit{CN}(0,1)$.

The indirect link consists of the BS-RIS and RIS-UE links. Given the fact that RIS is designed to be deployed on the top or surface of tall buildings, the BS-RIS link is assumed to be line-of-sight (LOS) transmission: 
\begin{equation}
 \label{eq2}
\bm{H}_{BR}=g_{BR}[\bm{h}_1,\bm{h}_2,...,\bm{h}_{N}]\in \mathbb{C}^{1\times N},
\end{equation}
where $N$ is the number of RIS elements, $g_{BR}$ is the path loss between BS and RIS, and $\bm{h}_{N} = \exp\left(\frac{{-2j\pi d_{N}}}{\lambda}\right) $ is the phase difference, herein $j=\sqrt{-1}$, $d_{N}$ is the distance between BS and RIS element $N$, $\lambda$ is the signal wavelength.

Then the RIS will reflect the signal to UEs via a phase shift vector $\bm{\theta}=[\theta_{1},\theta_{2},...,\theta_{N}]$, we define a diagonal matrix accordingly: 
\begin{equation}
 \label{eq4}
\bm{\Theta}=diag(\beta_{1}e^{j\theta_{1}},\beta_{2}e^{j\theta_{2}},...,\beta_{N}e^{j\theta_{N}})\in \mathbb{C}^{N\times N},
\end{equation}
where $\beta_{N}$ is the amplitude reflection coefficient and $\beta\in [0,1]$.

Considering the complex environment in the UE side, the RIS-UE link is presumed to be NLOS transmission, and we have      
\begin{equation}
 \label{eq5}
\bm{H}_{Rk}=g_{R,k}[\bm{h}'_{1,k},\bm{h}'_{2,k},...,\bm{h}'_{N,k}]\in \mathbb{C}^{1\times N},
\end{equation}
where $g_{R,k}$ is the path loss between RIS and UE $k$, and $h'_{N,k}$ is the small scale fading of the signal reflected by the RIS element $N$. Path loss is given by $g_{Rk}=d_{Rk}^{-{\alpha_{Rk}}/{2}}$, where $d_{Rk}$ is the distance from RIS to UE $k$, and $\alpha_{Rk}$ is the pathloss exponent.

Finally, the channel gain from BS $j$ to UE $k$ is:
\begin{equation}
 \label{eq6}
G_{j,k}=|\bm{H}_{B,k}+\bm{H}_{Rk}\bm{\Theta}\bm{H}_{BR}^{T}|^2.
\end{equation}
For the phase shift control, inspired by \cite{8888223}, we assume the channel state informations (CSIs) are perfectly shared between BS and RIS. The RIS phase shift is calculated by: $\theta_{N}=\arg(\bm{H}_{Bk})-\arg(g_{BR}h_{N}g_{R,k}h_{N,k}')$ to give every term in $\bm{H}_{Rk}\bm{\Theta}\bm{H}_{BR}^{T}$ the same phase as $\bm{H}_{Bk}$, then the total received signal will be strengthened. 

For a downlink transmission between BS $j$ and UE $k$, the transmission rate is:
\begin{equation}
\resizebox{0.9\hsize}{!}{$\begin{split}
C_{j,k}=&b_{j,k} \log_{2} \left(1+ \right.\\ & \left. \frac{\sum _{r\in{\mathcal{R}_{j,k}}} p_{j,r}a_{j,k,r}G_{j,k,r}}{b_{j,k}N_{0}+\sum\limits_{j'\in \mathcal{J}_{-j}}\sum\limits_{k'\in \mathcal{K}_{j'}}\sum\limits_{r'\in \mathcal{R}_{j'}}{p_{j',r'}a_{j',k',r'}G_{j',k',r'}}}\right),
\end{split}$}
 \label{eq7}
\end{equation}
where $\mathcal{R}_{j,k}$ is the set of resource blocks (RBs) allocated to UE $k$ by BS $j$\cite{3gpp}, $b_{j,k}$ is the total bandwidth allocated to UE $k$, $N_{0}$ is the power spectral density of noise, and $p_{j,r}$ is the transmission power of RB $r$ allocated by BS $j$. $a_{j,k,r}$ is a binary indicator. $a_{j,k,r}=1$ if the RB $r$ is allocated to UE $k$; otherwise $a_{j,k,r}=0$. $G_{j,k,r}$ denotes the channel gain between BS $j$ and UE $k$. $\mathcal{J}_{-j}$ denotes the set of BSs except BS $j$, $\mathcal{K}_{j'}$ is the UE set of BS $j'$, and $\mathcal{R}_{j'}$ is the RB set in BS $j'$. In this work, the RBs are allocated by the proportional fairness method\cite{b1}.  
\subsection{Energy Consumption Model}
The energy consumption model for the BS is:
\begin{equation} \label{eq8}
 P_{in}=\left\{
\begin{array}{lcl} P_{0}+\delta_{p} P_{out},     &      & 0 < P_{out} \leq P_{max}, \\
P_{sleep} ,    &      &  P_{out}=0,
\end{array} \right.
\end{equation}
where $P_{0}$ is the fixed power consumption, $\delta_{p}$ is the slope of load-dependent power consumption, $P_{out}$ is the transmission power, $P_{max}$ is the maximum transmission power, and $P_{sleep}$ is the constant power consumption in sleep mode \cite{b10}. 

\subsection{Problem Formulation}
The overall objective is to maximize the total EE, achieve the desired SINR for UEs, and prevent the BSs from overloading. Here the overloading means that current traffic demand has exceeded the transmission capability of one BS, and then the attached UEs may experience a long delay. The problem formulation is given by:
\begin{subequations}\label{e2:main}
\begin{align}
\max\limits_{P_{j}} \quad  & \frac{\sum_{j\in \mathcal{J} } \sum_{k \in \mathcal{K}_{j}} W_{j,k}}{\sum_{j\in \mathcal{J} } P_{j}}-\phi n_{od},  & \tag{\ref{e2:main}} \\
 \text{s.t.} \quad & (\ref{eq6})\, (\ref{eq7})\, (\ref{eq8}), & \label{e2:c}  \\
&\sum\limits_{k \in \mathcal{K}_{j}} \sum\limits_{r\in \mathcal{R}_{j}}{a_{j,k,r}} \leq |\mathcal{R}_{j}| ,& \label{e2:d}\\
&\sum\limits_{k \in \mathcal{K}_{j}} a_{j,k,r} \leq 1, & \label{e2:e} \\
&  SINR_{thr} \leq SINR_{j,k}, & \label{e2:f}
\end{align}
\end{subequations}
where $W_{j,k}$ is the throughput of UE $k$ in BS $j$, $P_{j}$ is the power consumption of BS $j$, and $n_{od}$ is the number of BSs that are overloaded.  We apply $\phi$ as a penalty factor to prevent overloading. Equation (\ref{e2:c}) is the system operation constraint, equation (\ref{e2:d}) indicates the number of available RBs can not exceed $|\mathcal{R}_{j}|$, equation (\ref{e2:e}) means one RB can only be allocated to at most one UE, and equation (\ref{e2:f}) is the SINR threshold constraint of UEs.

On one hand, turning off SBSs can greatly reduce energy consumption. But it will also increase the risk of MBS overloading, since the MBS has to take over the UEs of the sleeping small cells. Therefore, to maximize the total objective, we have to intelligently control the on/off status of SBSs to reduce the energy cost and overload risk, and following we will introduce an HRL based architecture.

\section{Hierarchical reinforcement learning for energy-efficient RAN}
\label{s4}
\subsection{Hierarchical Reinforcement Learning}

In traditional RL, the problem is defined by an MDP $<S,A,T,R>$, where $S$ is the set of states, $A$ is the set of actions, $T$ is the transition probability with $T:S\times A \times S$, and $R$ is the reward function. Then, one standalone agent will interact with the environment to maximize its long-term expected reward\cite{bzhou}. 

By contrast, in HRL, the agent consists of two controllers, namely meta-controller and sub-controller\cite{b7}. Accordingly, the MDP is rewritten by $<S,A,T,R,\mathcal{G}>$, where $\mathcal{G}$ indicates the set of goals. Based on current state $s\in S$, the meta-controller will generate high-level goals $g\in\mathcal{G}$ for sub-controllers. 
Then, these goals are transformed to high-level policies by the critic. Consequently, the sub-controller chooses low-level actions $a\in A$ according to high-level policies, and receives an intrinsic reward $r_{in}$. Finally, the meta-controller will receive an extrinsic reward $r_{ex}$ from the environment, and select new goals $g'$ for the sub-controller. The idea behind the HRL is to introduce hierarchy architecture in RL. In particular, the meta-controller will produce high-level policies to guide the low-level action selection of the sub-controller. Compared with traditional RL, HRL is considered as a more efficient learning method due to the hierarchical architecture, and by dividing sub-goals it allows better management of multiple functionalities in RAN.    

\subsection{MDP Definition}
To transform the problem formulation showed by equation (\ref{e2:main}) into the HRL notation, the following MDP for sub-controllers and meta-controller are defined. 

Each SBS is regarded as a sub-controller, the MDP is defined by: 
\begin{itemize}
    \item \textbf{State}: The state $s_{sub}$ of SBS $j$ is defined by its traffic load ratio $s_{sub}=\{d_{SBS}\}$, which is given by:  
    \begin{equation}
    d_{SBS}=\frac{\sum_{k \in \mathcal{K}_{j}} D_{j,k}}{D^{max}_{j}} , \label{eq9}  
    \end{equation}
    where $\mathcal{K}_{j}$ indicates the set of UEs that are served by SBS $j$, $D_{j,k}$ is the traffic demand of UE $k$. $D^{max}_{j}$ is the max traffic load of SBS $j$, which is considered as a constant value to normalize the current traffic load of SBS. Meanwhile, note that the transmission demand of UEs often shows strong statistical regularity, and we assume the daily traffic load follows the patterns in \cite{b5}.
    \item \textbf{Action:} Based on $s_{sub}$, the SBS may change its transmission power $P_{SBS}$ to adapt the traffic demand. Then, the action is defined by $a_{sub}=\{P_{SBS}\}$.
    
    \item \textbf{Intrinsic reward:} The intrinsic reward of SBS is:
    \begin{equation}\label{eq10}
    r_{in}=\frac{\sum_{k \in \mathcal{K}_{j}} W_{j,k}}{P_{SBS}}-\phi n_{od},
    \end{equation}
    where $W_{j,k}$, $\phi$ and $n_{od}$ have been defined in equation (\ref{e2:main}). $r_{in}$ aims at maximizing its own EE and preventing overloading.
\end{itemize}

The meta-controller is responsible for the high-level policies for the agent. The MBS is defined as the meta-controller, and its MDPs are:
\begin{itemize}
   \item \textbf{State}: The state of meta-controller consists of the traffic load ratio of SBSs:
    \begin{equation}
    s_{meta}=\{d_{SBS,j}\}, j\in \mathcal{J}_{SBS},  \label{eq11}  
    \end{equation}
    where $d_{SBS,j}$ is the load ratio of SBS $j$, and $\mathcal{J}_{SBS}$ is the set of SBSs. 
    
    \item \textbf{Goals for sub-controller}: With the traffic load status of the SBSs, MBS can generate high-level policies for the SBSs. The goals $g_{meta}$ are turning on/off the SBSs:
    \begin{equation}
    g_{meta}=\{q_{SBS,j}\}, j\in \mathcal{J}_{SBS},  \label{eq12}  
    \end{equation}
    where $q_{SBS,j}$ is a binary variable to indicate the on/off status of SBS $j$. $q_{SBS,j}=1$ means keeping the SBS $j$ active, otherwise $q_{SBS,j}=0$ denotes turning off SBS $j$ to save energy.   
    
   \item \textbf{Extrinsic reward}: The meta-controller focuses more on the overall performance of the whole cell. Accordingly, the extrinsic reward is given by the objective of the problem formulation in equation (\ref{e2:main}):
   \begin{equation}\label{eq13}
     r_{ex}=\frac{\sum_{j\in \mathcal{J} } \sum_{k \in \mathcal{K}_{j}} W_{j,k}}{\sum_{j\in \mathcal{J} } P_{j}}-\phi n_{od},
   \end{equation}

\end{itemize}

\subsection{Q-value Update and Goal Selections}

In this section, we introduce how to update the Q-values of controllers,  and the action and goal selection strategies.

The Q-values of meta-controller is updated by:
\begin{equation} \label{eq14}
\begin{aligned}
&Q_{meta}^{new}(s_{meta},g_{meta}) = Q_{meta}^{old}(s_{meta},g_{meta})+\\
&\alpha(r_{ex}+\gamma \max\limits_{g} Q_{meta}(s_{meta}',g)-Q_{meta}^{old}(s_{meta},g_{meta})),
\end{aligned}
\end{equation}
where $s_{meta}'$ denotes the next state, $\alpha$ is the learning rate, and $\gamma$ is the discount function ($0< \alpha, \gamma <1$). $Q^{old}_{meta}$ and $Q^{new}_{meta}$ denote old and new Q-values for meta-controller, which means the accumulated reward brought by state-goal pair $(s_{meta},g_{meta})$. Then we use the $\epsilon$-greedy policy for goal selection:
\begin{equation} \label{eq15}
\pi(s_{meta}) =\left\{
\begin{array}{ccl}
 \arg~\max\limits_g \, Q(s_{meta},g) ,  &  rand>\epsilon,\\
  \text{random~goal~selection}, &   rand\leq\epsilon.\\
\end{array} \right.
\end{equation}
where $rand$ is a random number between 0 and 1, and $\epsilon <1$. $\epsilon$-greedy policy can balance the exploration and exploitation of goals to maximize the long-term reward.

Similarly, for the sub-controller, the Q-values are updated:
\begin{equation} \label{eq16}
\resizebox{1\hsize}{!}{$\begin{aligned}
Q&_{sub}^{new}(s_{sub},g_{meta},a_{sub}) = Q_{sub}^{old}(s_{sub},g_{meta},a_{sub})+\\
&\alpha(r_{in}+\gamma \max\limits_{a} Q_{sub}(s_{sub}',g_{meta}',a)-Q_{sub}^{old}(s_{sub},g_{meta},a_{sub})),
\end{aligned}$}
\end{equation}
where $s_{sub}'$ is the next state, $g_{meta}'$ is the next goal generated by meta-controller,  $Q^{new}_{sub}$ and $Q^{old}_{sub}$ are new and old Q-values for sub-controller, respectively.
We still use $\epsilon$-greedy policy for the action selection of sub-controller: 
\begin{equation} \label{eq17}
\pi(s_{sub}) =\left\{
\begin{array}{ccl}
 \arg~\max\limits_a \, Q(s_{sub},g_{meta},a) ,  &  rand>\epsilon,\\
  \text{random~action~selection}, &   rand\leq\epsilon.\\
\end{array} \right.
\end{equation}

The HRL based sleep and transmission power control is summarized in Algorithm 1.

\begin{algorithm}h
	\caption{HRL algorithm for SBS sleep and power control}
	\begin{algorithmic}[1]
		\STATE \textbf{Initialize:} Wireless network and HRL parameters.
		\FOR{$episode$=1 to $Total$ }
		\FOR{MBS}
		\STATE With probability $\epsilon$ choose goals randomly, otherwise select $g_{meta}$ by $\arg~\max\limits_g Q(s_{meta},g)$ (Shown by equation (\ref{eq15})).
		\FOR{Each active SBS}
		\STATE With probability $\epsilon$ choose $a_{sub}$ randomly, otherwise select $a_{sub}$ by $\arg~\max\limits_a Q(s_{sub},g_{meta},a)$ (Shown by equation (\ref{eq17})).     
		\STATE Calculating intrinsic reward $r_{in}$, updating state $s_{sub}$ and Q-values by equation (\ref{eq16}).
		\ENDFOR
		\STATE MBS calculates extrinsic reward $r_{ex}$, updating state $s_{meta}$ and Q-values by equation (\ref{eq14}).  
		\ENDFOR
		\ENDFOR
		\STATE \textbf{Output:} Optimal SBS sleep and transmission power control strategy. 
	\end{algorithmic}
\end{algorithm}

\section{Performance Evaluation}
\subsection{Simulation Settings}
\begin{figure}[!b]
\centering
\vspace{-10pt}
\setlength{\abovecaptionskip}{-5pt} 
\includegraphics[width=6.4cm,height=4.3cm]{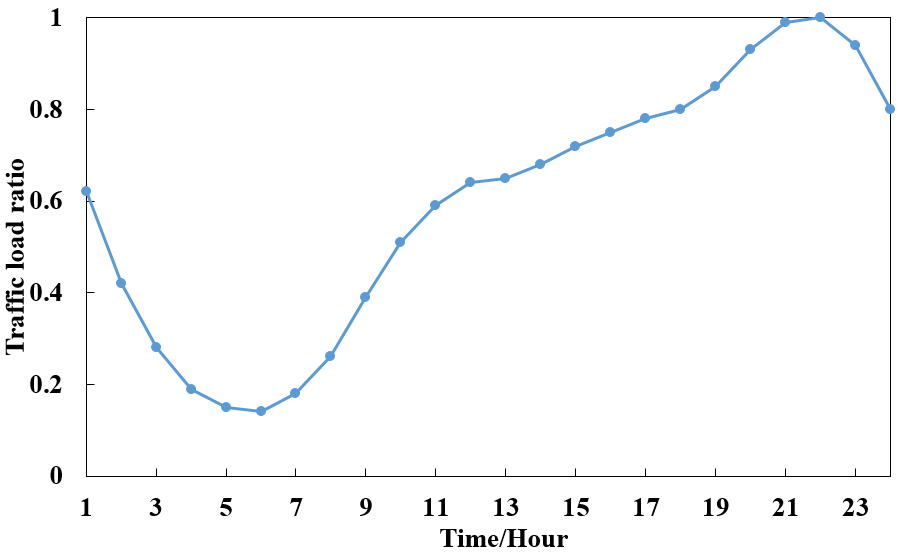}
\caption{Daily traffic load pattern of residential area.}
\label{fig2-1}
\end{figure}

In the simulations, we consider a dense urban environment in the MATLAB simulation platform, where there are 4 SBSs and 4 RISs. The coverage radius of MBS and SBS are 400m and 80m, respectively. The cell includes 50 randomly distributed UEs. The fixed power consumption of MBS and SBS are 130W and 75W, and the load-dependent power consumption slope is 4.7 and 2.6 for MBS and SBS, respectively \cite{b10}. We assume a deep sleep mode at the SBS with 0 power consumption. Each RIS has 10 reflecting elements with $3$ bits phase shift resolution. We assume the RIS power consumption is very low and it is not included in the power consumption. 
The path loss exponent for LOS and NLOS are 2.5 and 3.5, respectively \cite{b4}. The available bandwidth for each BS is $b_R = 20$ MHz. The traffic pattern is presumed to follow Fig. \ref{fig2-1}, which is a typical residential area traffic pattern \cite{b5}. The initial learning rate is 0.95, and we decay the learning rate after every several episodes for a stable learning performance, and the discount factor is 0.3. The simulation is repeated for 10 runs in MATLAB, and we present the average results with 95\% confidence interval.

\subsection{Simulation Results}

In this section, we include 4 cases: (1) no RIS and no sleep control (typical-cell), (2) sleep control without RIS (sleep-only), (3) RIS without sleep control (RIS-only), and (4) combining RIS with sleep control (RIS-sleep). We apply conventional Q-learning for case (1) to (3), and HRL for our proposed case (4).

\begin{figure}[!t]
\centering 
\setlength{\abovecaptionskip}{-2pt} 
\includegraphics[width=7cm,height=4.7cm]{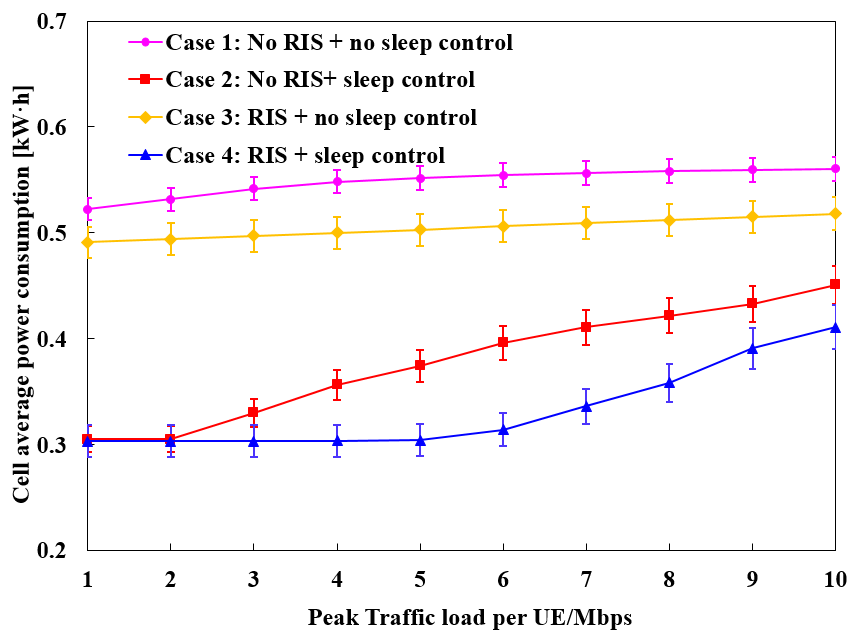}
\caption{Total power comparison of all BSs against peak traffic load.}
\label{fig3}
\vspace{-10pt}
\end{figure}

\begin{figure}[!t]
\centering 
\setlength{\abovecaptionskip}{-2pt} 
\includegraphics[width=7cm,height=4.7cm]{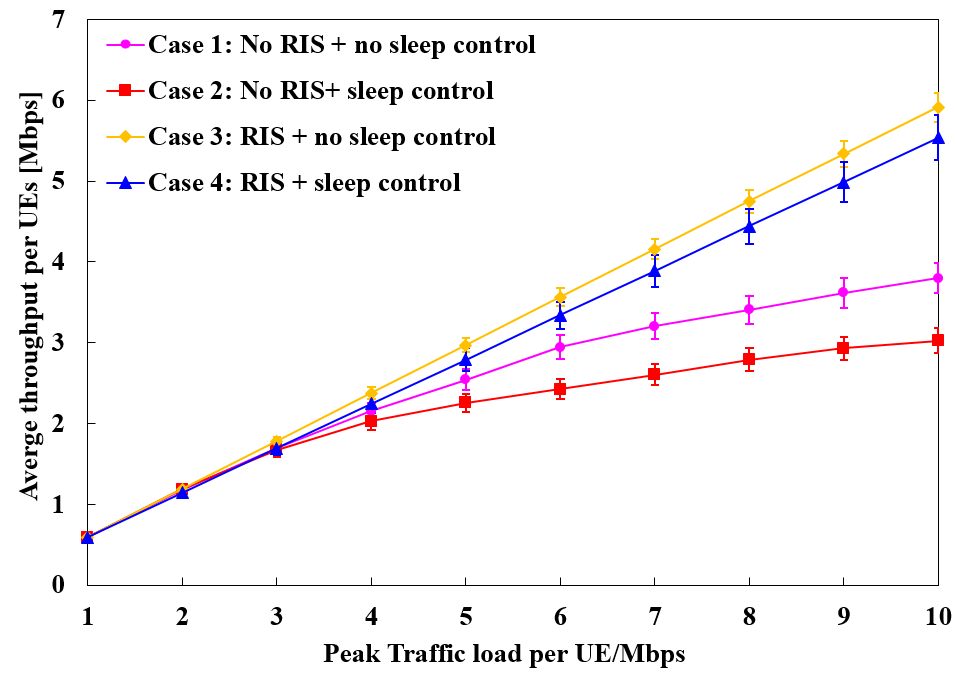}
\caption{Average throughput per UE in the cell against peak traffic load.}
\label{fig4}
\vspace{-10pt}
\end{figure}

\begin{figure}[!t]
\centering 
\setlength{\abovecaptionskip}{-2pt} 
\includegraphics[width=7cm,height=4.7cm]{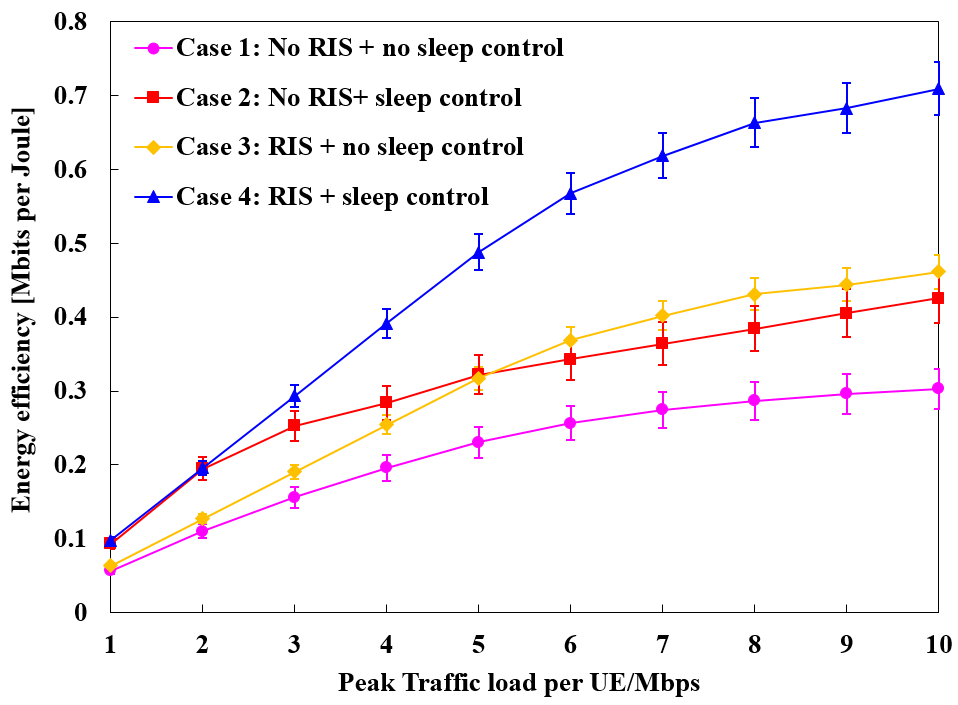}
\caption{EE of the BSs against peak traffic load.}
\label{fig5}
\vspace{-10pt}
\end{figure}

Fig. \ref{fig3} to \ref{fig5} first present the total power consumption of the BSs, average throughput per UEs, and EE against peak traffic load for the 4 cases, respectively. One can observe that: (i) typical-cell, as a benchmark here, presents the highest power consumption and lowest EE; (ii) comparison between typical-cell and sleep-only in Fig. \ref{fig3} demonstrates that sleep control can significantly reduce the power consumption; (iii) the EE results of typical-cell and RIS-only in Fig. \ref{fig4} shows that RIS is highly beneficial to the average throughput. 

More specifically, as shown in Fig. \ref{fig4}, when the traffic load is lower than 4 Mbps, the existing channel capacity is already huge enough to serve the UEs. However, when the peak traffic load becomes higher than 5 Mbps, RIS-only and RIS-sleep show a higher throughput than other two cases, which can be explained by RIS's capability to improve the SINR of UEs. 

When it comes to the EE metric, as shown by Fig. \ref{fig5}, the case 4, namely RIS-sleep strategy, displays the best EE performance. On the contrary, typical-cell shows the worst EE performance due to the absence of both RIS and sleep control. sleep-only and RIS-only have comparable EE. It is observed that the former strategy has lower power consumption and lower throughput, and RIS-only is the opposite (indicated by Figs. \ref{fig3} and \ref{fig4}). As a result, these two cases show a close EE. When the peak traffic load is 8 Mbps, RIS-sleep achieves a more than doubled EE than other cases. 

\begin{figure}[!t]
\centering
\setlength{\abovecaptionskip}{-2pt} 
\includegraphics[width=7cm,height=4.7cm]{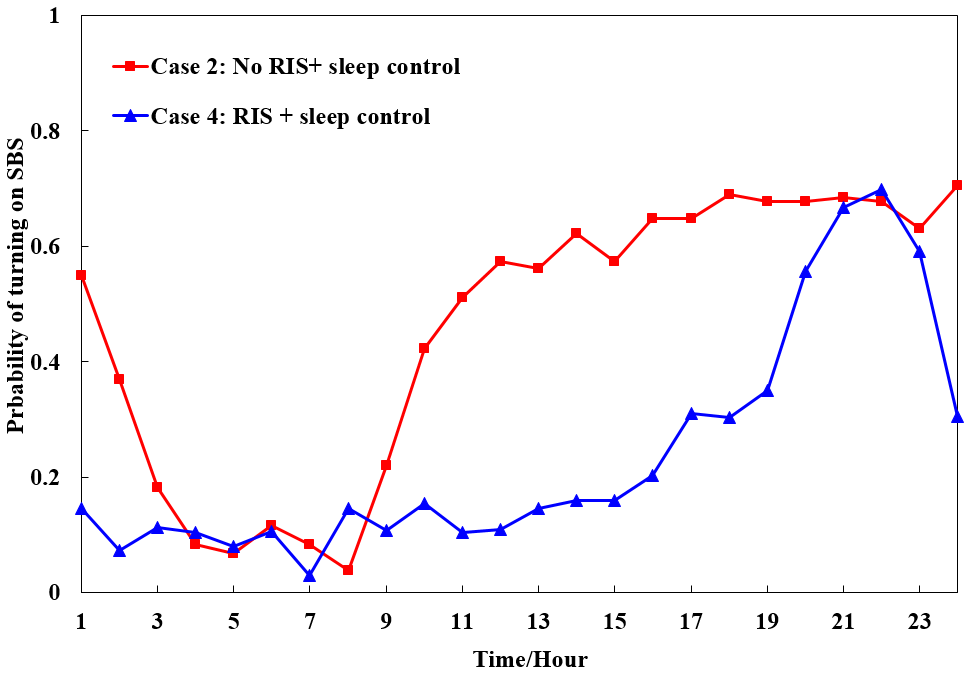}
\caption{Probability of keeping SBSs active under 8 Mbps peak traffic load.}
\label{fig6}
\vspace{-10pt}
\end{figure}

To better explain how RIS and sleep control are combined, sleep-only and RIS-sleep are compared in Fig. \ref{fig6} in terms of the possibility of keeping SBSs active. During the off-peak period (from 3:00 to 9:00 in the traffic patterns shown by Fig. \ref{fig2-1}), most SBSs are shut off to save energy, and the existing traffic demand is served by MBS. However, after 11:00, sleep-only has to turn on most SBSs to satisfy the increasing traffic load, otherwise the MBS will be overloaded and the total throughput will be greatly affected. By contrast, RIS-sleep is capable of keeping most SBSs sleep until 17:00, because MBS can process the increasing traffic demand with a higher SINR provided by RIS.

\begin{figure}[!t]
\centering
\setlength{\abovecaptionskip}{-2pt} 
\includegraphics[width=7cm,height=4.5cm]{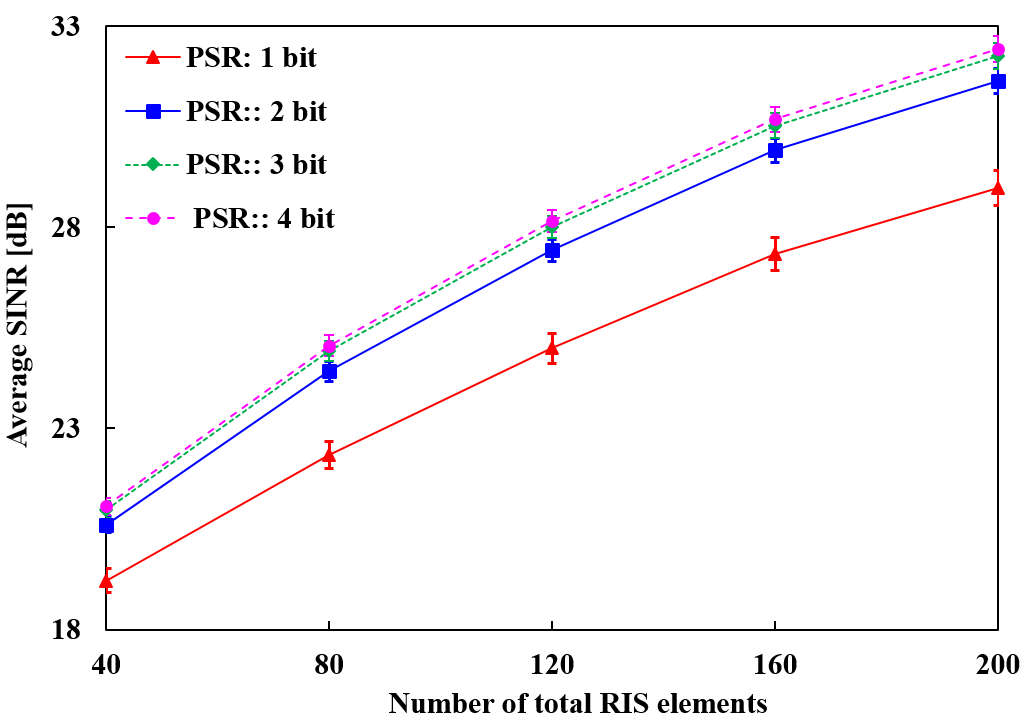}
\caption{Average SINR of UEs under different phase shift resolutions.}
\label{fig7}
\vspace{-10pt}
\end{figure}

Apart from the aforementioned discussions, we further investigate the average SINR of UEs against the number of RIS reflecting elements in Fig. \ref{fig7} under different RIS phase shift resolutions (PSR). A higher PSR generally indicates a more accurate phase shift design. One can observe that more RIS elements and higher PSR are as expected essentially useful to improve the SINR of UEs. On the other hand, the improvements brought by PSR are barely observable from 3 to 4 bits. 

\begin{figure}[!t]
\centering
\setlength{\abovecaptionskip}{-2pt} 
\includegraphics[width=7.6cm,height=4.5cm]{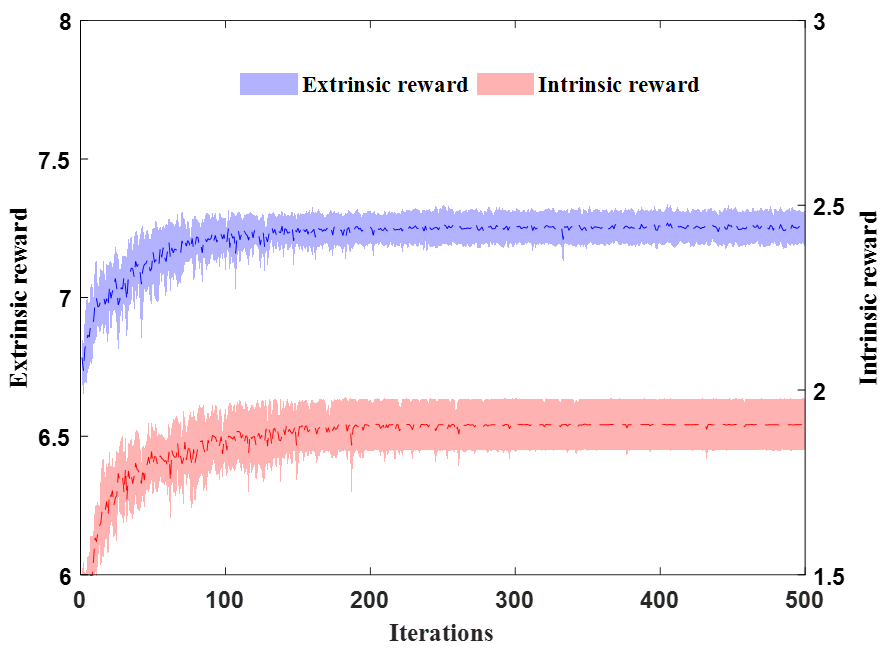}
\caption{Convergence performance analyses.}
\label{fig8}
\vspace{-10pt}
\end{figure}

Finally, Fig. \ref{fig8} presents the convergence performance. It shows that both intrinsic and extrinsic rewards increase with more iterations and finally converge, which means that meta controller and sub-controller are well coordinated to maintain the overall performance.

\section{Conclusion}
The reconfigurable intelligent surface is a promising technology to enable 5G beyond and 6G networks. In this paper, we combine reconfigurable intelligent surfaces with sleep control to improve the energy efficiency of heterogeneous 5G radio access networks. We propose a hierarchical reinforcement learning-based method to optimize the sleep control strategy of small base stations. Compared with the standalone sleep control method, the simulations show a significantly higher energy efficiency by jointly deploying reconfigurable intelligent surface and sleep control in a hierarchical learning framework. In addition, we conclude that (i) sleep control largely contributes to reducing power consumption and improving energy efficiency; (ii) reconfigurable intelligent surface is beneficial to the average throughput, especially for high traffic load conditions. In the future, we will investigate the control strategy of the phase shift of reconfigurable intelligent surfaces.

\section*{Acknowledgement}
This work has been supported by MITACS and Ericsson Canada, and NSERC Collaborative Research and Training Experience Program (CREATE) under Grant 497981.

\bibliographystyle{IEEEtran}
\bibliography{Globecom2022}

\end{document}